# Diatom-inspired architected materials using language-based deep learning: Perception, transformation and manufacturing

Markus J. Buehler[*,#]

[*]Laboratory for Atomistic and Molecular Mechanics, Massachusetts Institute of Technology, 77 Massachusetts Avenue, Cambridge, MA, 02139, USA, mbuehler@MIT.EDU

**ABSTRACT**

Learning from nature has been a quest of humanity for millennia. While this has taken the form of humans assessing natural designs such as bones, butterfly wings, or spider webs, we can now achieve generating designs using advanced computational algorithms. In this paper we report novel biologically inspired designs of diatom structures, enabled using transformer neural networks, using natural language models to learn, process and transfer insights across manifestations. We illustrate a series of novel diatom-based designs and also report a manufactured specimen, created using additive manufacturing. The method applied here could be expanded to focus on other biological design cues, implement a systematic optimization to meet certain design targets, and include a hybrid set of material design sets.

## 1. INTRODUCTION

Bioinspiration has been an active field of research. Among its pioneers is David Taylor – the inaugural editor of Journal of the Mechanical Behavior of Biomedical Materials and a fixture at the regular International Conference on Mechanics of Biomaterials & Tissues. Professor Taylor has contributed to the biomaterials community in so many ways, including his dedication to building a strong community around the journal and the conference, from which a wide set of research contributions have spawned. To honor David Taylor, this article is dedicated to him in celebration of his retirement (Palomba et al., 2020; Taylor, 2007, 2008; Taylor et al., 2007).

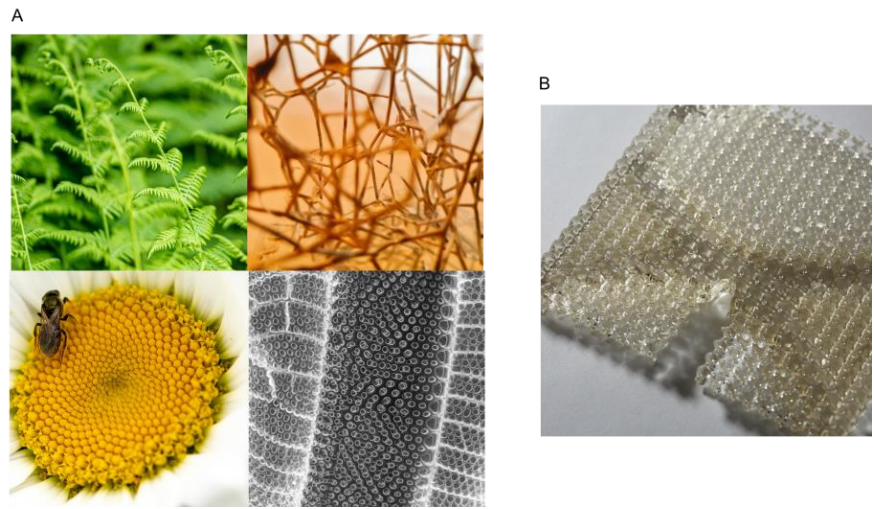

**Figure 1**: Hierarchical structures in natural materials (A), e.g. fern leaves, spider web, patterns in flowers, diatom structure (diatom courtesy Picturepest - Diatom - Isthmia nervosa - 400x, CC BY 2.0, https://commons.wikimedia.org/w/index.php?curid=39164600). B, Biologically inspired material design, where a gyroid material architecture is modulated to mitigate the effects of a notch, minimizing stress concentrations.

While bio-inspired design has taken the form of humans assessing natural designs such as bones, butterfly wings, plans structures, or spider webs, and many others (some examples shown in **Figure 1**), we can now generate designs using advanced computational algorithms (Buehler, 2022a; Giesa et al., 2012; Hsu et al., 2022; Spivak et al., 2011a, 2011b; Yang & Buehler, 2021). In this paper we report novel biologically inspired designs of diatom structures, enabled using transformer neural networks, using natural language models to learn, process and transfer insights across manifestations.

The approach used in this paper follows a similar strategy as reported in (Hsu et al., 2022; Yang & Buehler, 2021). However, for the present study we train a VQGAN model from scratch, similar as done in (Buehler, 2022b), using a publicly available diatom dataset (ADIAC Project (CEC Contract MAS3-CT97-01). The examples provided here offer a perspective to consider bio-inspiration from an artificial intelligence point of view; and the integration of human language as design input exemplifies how artificial and human intelligence can work together towards innovative design solutions.

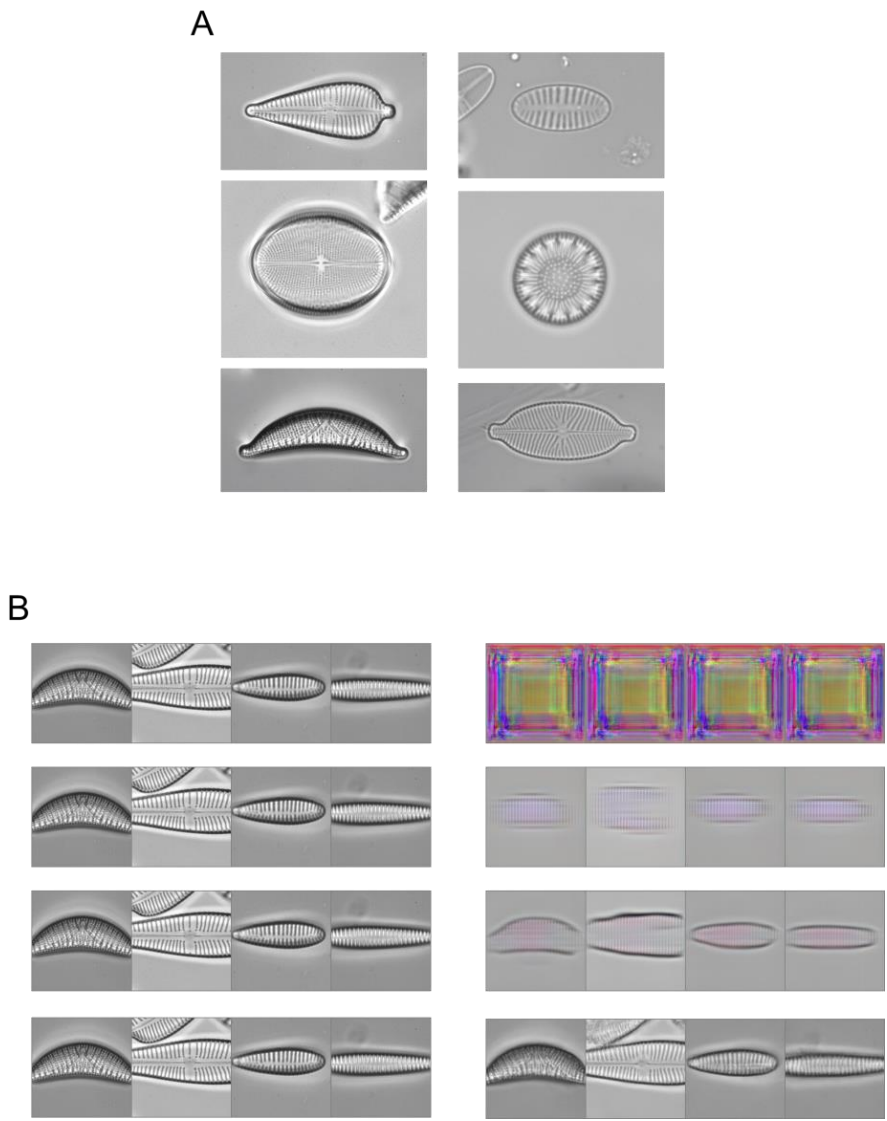

**Figure 2**: A, samples of diatom structures as reported in (ADIAC Project (CEC Contract MAS3-CT97-01)).. B, training of a VQGAN model, so that it can reproduce the particular architectural details of diatom structures. At the end of the training, the transformer neural network has learned how to generate synthetic diatom structures. The samples shown here (left: ground truth, right: reconstruction) show how performance increases over a total of 200 training epochs.

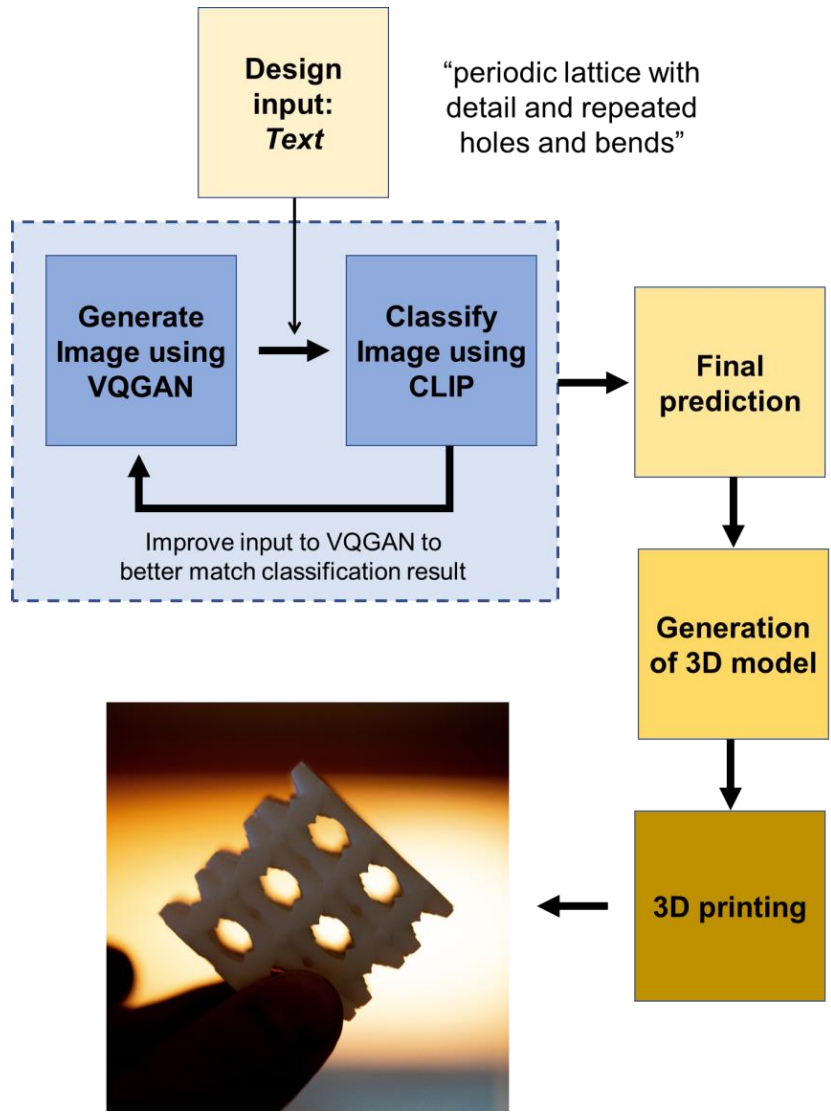

**Figure 3**: Overall approach used here, following the method reported in (Hsu et al., 2022; Yang & Buehler, 2021). Design inputs are provided via a text prompt, which the algorithm converts into a final image prediction. It is then used to generate a 3D model through a tilation and 2D-3D translation algorithm, resulting in a final material manufactured using 3D printing.

The focus of this chapter is on diatoms, given that they provide interesting multifunctional material designs. A challenge has been to translate the interesting structural features seen in diatoms into engineering solutions. One way is to extract salient features and then reconstruct bioinspired analogues; this method uses human intelligence to achieve the goal and has been widely used. Here we propose a complementary method that removes some of the human biases, and gives an artificial intelligence system the task to take design cues from a library of diatom structures and covert them into a set of design solutions.

## 1.1   Training the transformer image generation model

The first step in the approach used here is to train a neural network that has learned to generate biological designs. In the study reported here, we develop such a model by training it against a set of diatom structures as reported in (ADIAC Project (CEC Contract MAS3-CT97-01) (**Figure 1**). The data is then used for training a VQGAN model (Esser et al., 2020), so that it can reproduce the particular architectural details of diatom structures. At the end of the training, the transformer neural network has learned how to generate synthetic diatom structures. The samples shown in **Figure 1B** (left: ground truth, right: reconstruction) show how performance increases over a total of 200 training epochs.

## 1.2 Integrating the transformer model with CLIP

We now integrate the image generation algorithm with a classification method, CLIP. CLIP is a general-purpose image classifier, as reported in (Radford et al., 2021). VQGAN and CLIP is integrated as suggested in (Crowson et al., 2022).

**Figure 3** depicts the overall approach similar as done in (Hsu et al., 2022; Yang & Buehler, 2021). Design inputs are provided via a text prompt, which the algorithm converts into a final image prediction. It is then used to generate a 3D model through a tileation and 2D-3D translation algorithm, resulting in a final material manufactured using 3D printing.

**Figure 4** shows sample predictions using the trained model, for a variety of design inputs. As can be seen, a variety of structures are generated that feature aspects of diatoms, but also accomplish to represent aspects of the design cues provided. This integration of biological data, human language, and resulting designs that can be assessed and examined, offers a new approach to bio-inspiration.

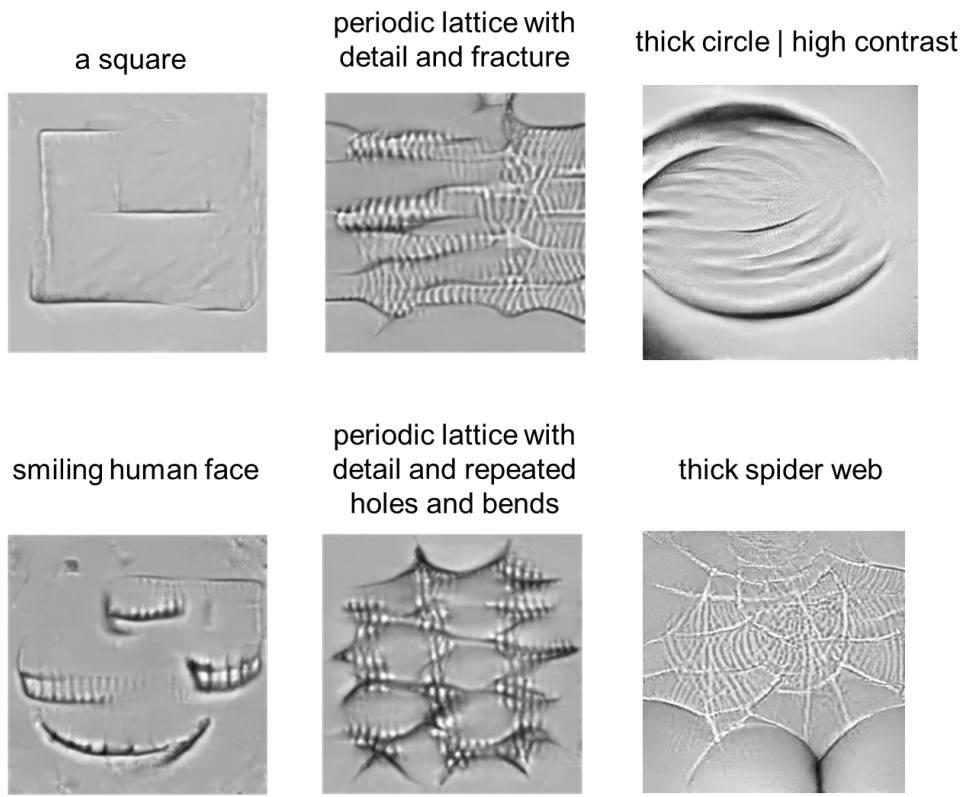

**Figure 4**: Sample predictions using the trained model, for a variety of design inputs. The resulting images reflect characteristic features of diatoms, but also meet the design demands given by the text prompt.

## 2. RESULTS AND DISCUSSION

We now present several examples of using this model to generate diatom-inspired materials that are manufactured using 3D printing. We start with a first example, shown in **Figure 5**, where we synthesize high-resolution images that are converted into height maps for 3D printing. This is the simplest way by which a 2D image can be transformed into a 3D material. **Figure 5** shows results from two sample text prompts, including the analysis and manufactured material.

Next, we move to generate and manufacture full 3D architected materials, using the algorithm reported in (Hsu et al., 2022). **Figure 6** shows Generation of 3D architected materials, using the algorithm reported in (Hsu et al., 2022). **Figure 6A** depicts the progression from the text prompt, to the generation of a processed image, and the translation to a periodic 3D structure. **Figure 6B** illustrates the manufacturing method. **Figure 6C** shows results of a finite element analysis of the structure, exposed to tensile loading.

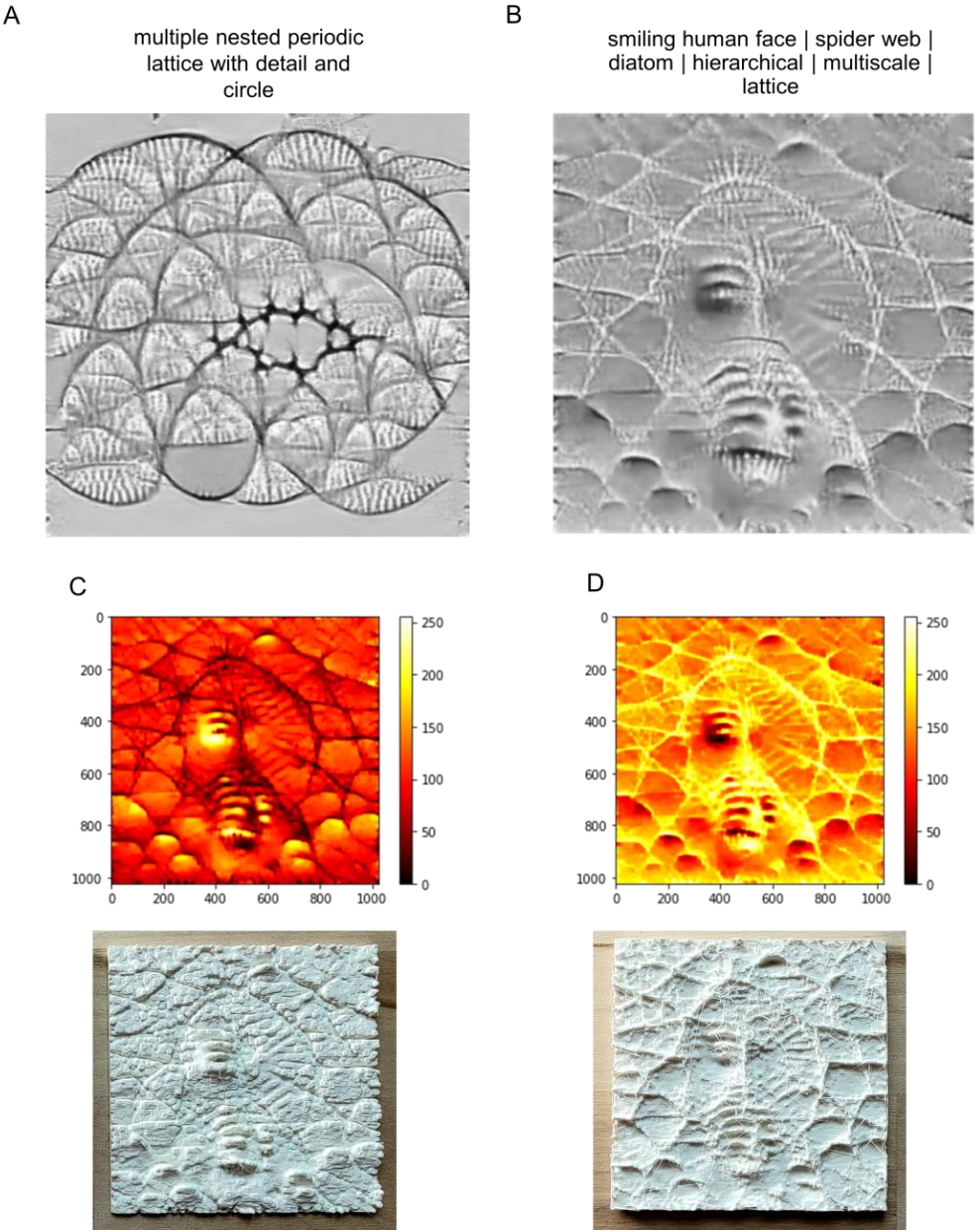

**Figure 5**: Additional examples, using higher resolution image output (1024x1024), which can yield results with intricate details. Panels A-B show two sample text prompts, with panels C-D showing two distinct post-processing methods. In C, a contour height map is created where dark areas are correlated with high height. In D, the inverse is achieved, where bright areas are associated with high height, forming a sort of inverse design to what is shown in C. The bottom rows show the 3D printed results (7cm x 7cm size; printed using FDM with a Ultimaker S3 printer using PLA filament).

Such analysis can provide insights into areas of large displacements, high stresses, or other mechanical measures. The finite element analysis is conducted using an isotropic material model where the material parameters are chosen to match the datasheet of the Ultimaker PLA filament.

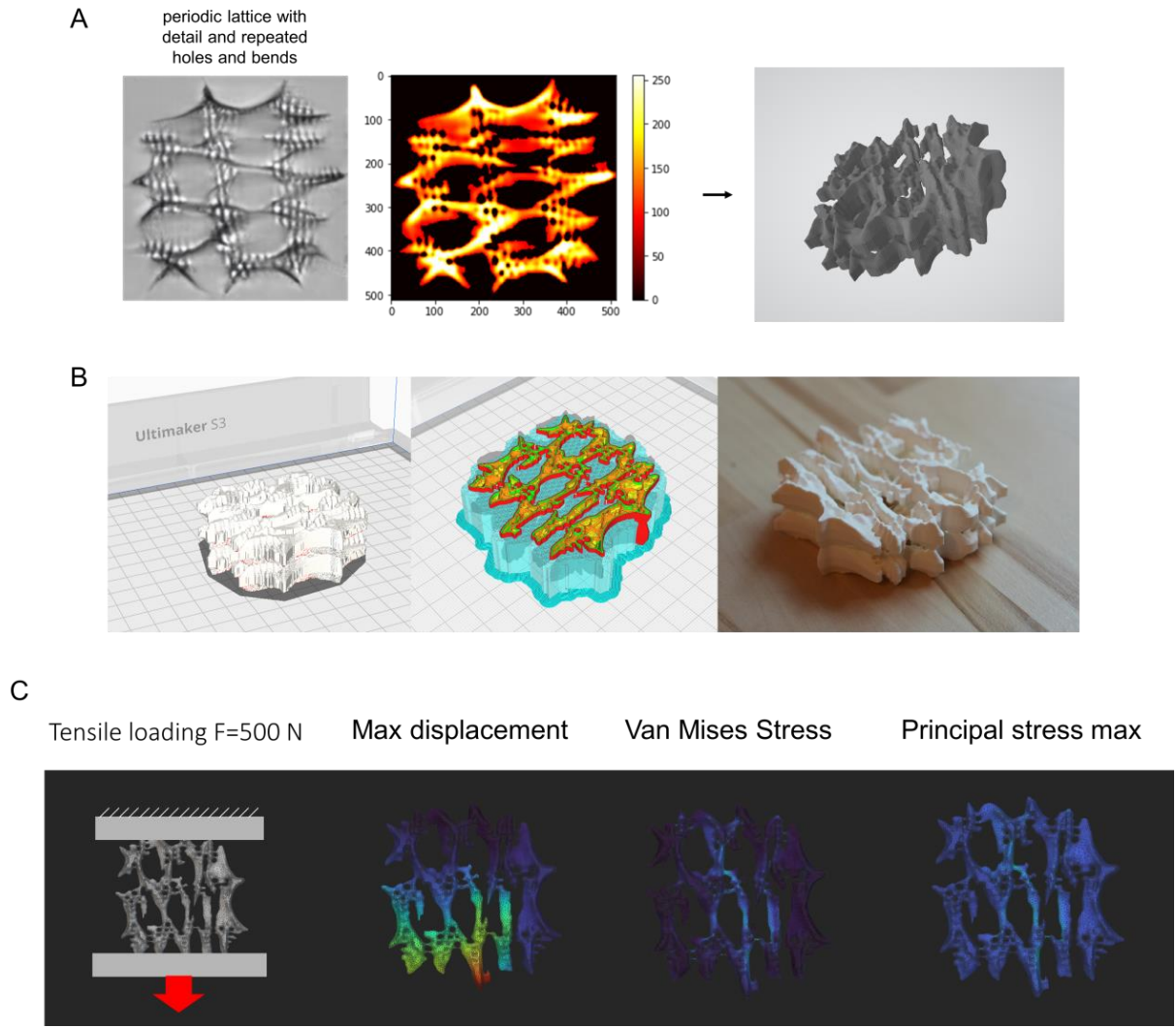

**Figure 6**: Generation of 3D architected materials, using the algorithm reported in (Hsu et al., 2022). Panel A shows the progression from the text prompt, to the generation of a processed image, and the translation to a periodic 3D structure. Panel B shows the manufacturing process by which the material is made experimentally (cross-dimensions 7cm x 7 cm; printed using FDM with a Ultimaker S3 printer using PLA filament and PVA as support material to realize the complex 3D geometry). Panel C shows a finite element analysis of the structure (performed using a static analysis via the *nTopology* software, exposed to tensile loading. Such analysis can provide insights into areas of large displacements, high stresses, or other mechanical measures.

The design reported in **Figure 6** reflects a form of a synthetic diatom. However, it does not yet possess internal structure. One way to add another level of hierarchical structuring is to use the stress field resulting from the finite element analysis shown in **Figure 6C** as an input to modulate a porous gyroid microstructure**. Figure 7** shows results of generation of such a multi-level architected material. **Figure 7A** shows how the Van Mises stress distribution is converted into a field map, which in turn is used to modulate the thickness of a gyroid microstructure. Areas of high stress yield solid material, and areas of low stress a highly porous material; the cross-sections directly visualize the multi-level structure obtained in this way. **Figure 7B** shows the results of manufacturing of the resulting material design using resin printing. **Figure 7C** shows images of the resulting material with multilevel material architecture; left: overall specimen (scale bar 1 cm), middle:

different perspective against background light to visualize the internal structure. Right: Macro-view of the internal gyroid structure in an area of high porosity (scale bar: 5 mm).

**Figure 8** shows another example of using a simpler design as an elementary unit cell, rendered fully periodic and tileable in 3D to generate a block of architected material (4cm x 4cm x 4cm). **Figure 8A**, left shows the design cues, the unit cell (middle), and the resulting design as an architected material cube. **Figure 8B** offers different views of the designed material and **Figure 8C** reveals the manufactured result, showing the intricate internal structure.

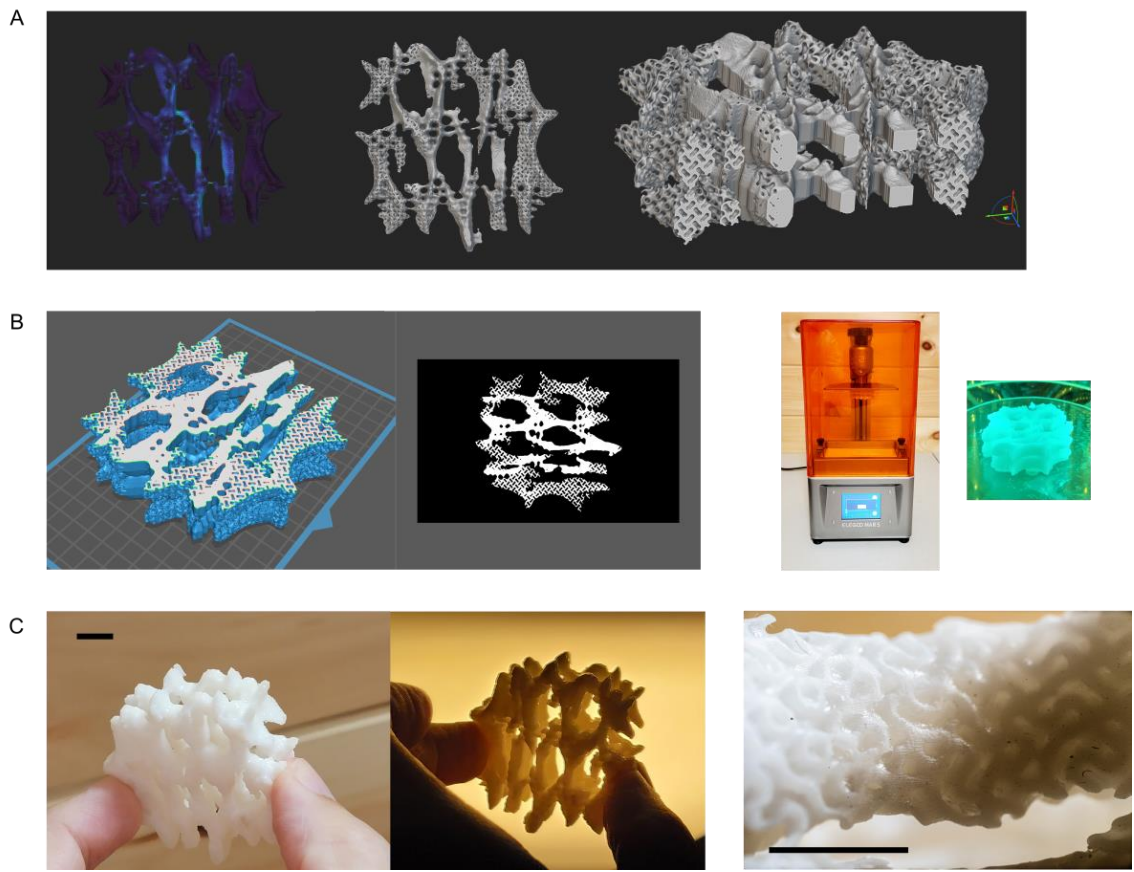

**Figure 7**: Generation of multi-level architected material. A shows how the Van Mises stress distribution is converted into a field map, which in turn is used to modulate the thickness of a gyroid microstructure. Areas of high stress yield solid material, and areas of low stress a highly porous material. The cross-section (A, right) shows this internal microstructure distribution. B, Manufacturing of the resulting material design using resin printing (ABS-like resin; printed using a Elegoo Mars printer and subsequently cured in a UV bath). C, Resulting material with multilevel material architecture; left: overall specimen (scale bar 1 cm), middle: different perspective against background light to visualize the internal structure. Right: Macro-view of the internal gyroid structure in an area of high porosity (scale bar: 5 mm).

## 4. CONCLUSION

In this chapter, a method to translate biological structural data – here, diatoms, is applied to generate various biologically inspired designs. The various results showed that natural materials provide a rich set of inspiration and how 3D printing (both resin-based and fused-deposition modelling (FDM)) can be powerful tools to manufacture resulting designs.

The method applied here could be expanded to focus on other biological design cues, implement a systematic optimization to meet certain design targets, and include a hybrid set of material design sets.

Future work could also focus on testing some of the resulting materials, and compare against simulation results. An area of particular interest could be to test fracture properties to better understand whether the porous geometry of diatoms (Garcia, Pugno, et al., 2011; Garcia, Sen, et al., 2011) realized through this deep learning enabled translational approach can yield interesting new properties.

The intersection of artificial and human intelligence is an exciting frontier that offers a platform for novel solutions; especially, converging towards new methods that systematically mine natural material data for incorporation into synthetic design solutions. Not only can such work advance our fundamental understanding of biological design principles, but also further our capacity to combine knowledge in existing engineering theories with evolutionary concepts.

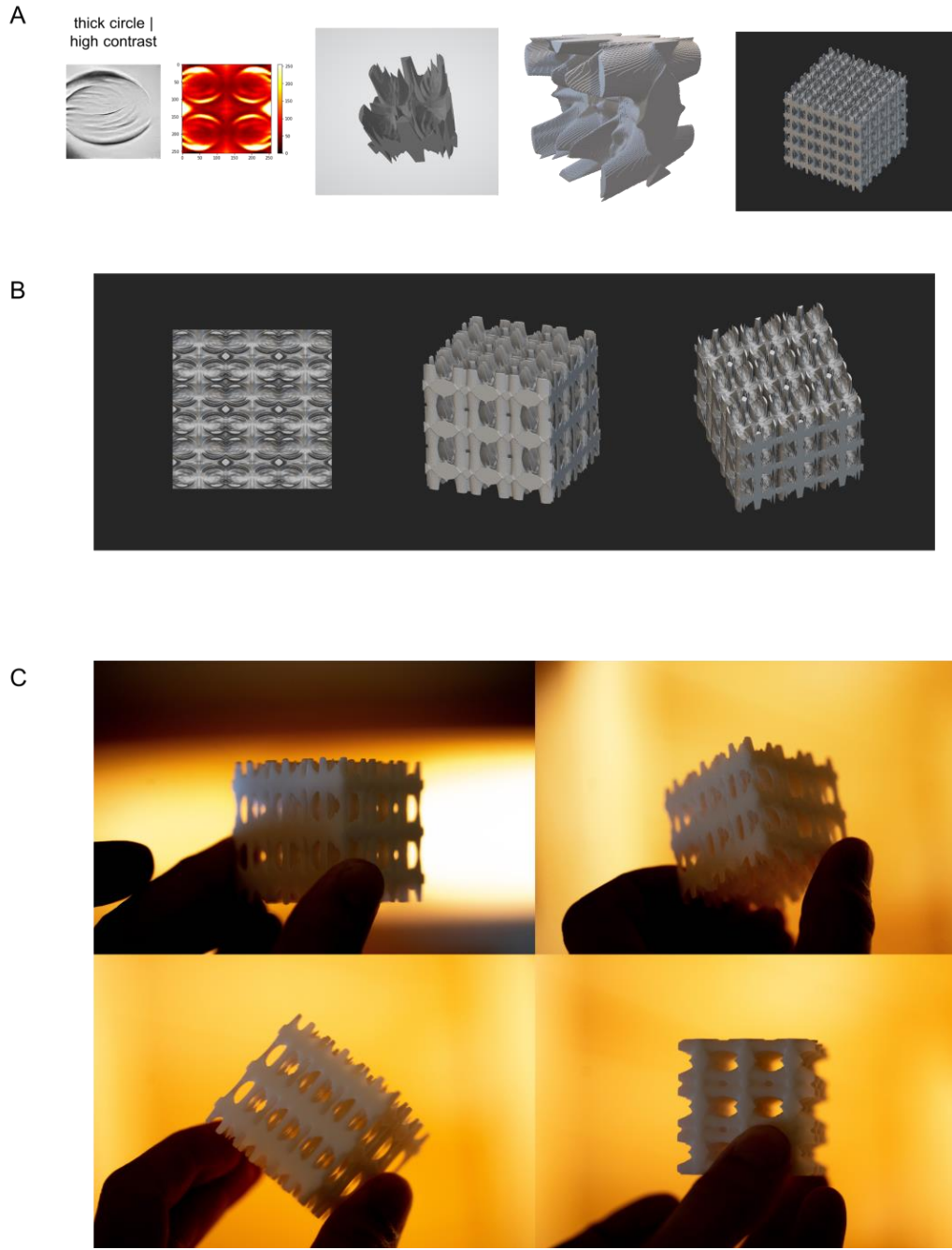

**Figure 8**: Another example of using a simpler design as an elementary unit cell, rendered fully periodic and tileable in 3D to generate a block of architected material (4cm x 4cm x 4cm). A, left shows the design cues, the unit cell (middle), and the resulting design as an architected material cube. B, Different views of the designed material. C, Manufactured result, from different angles, showing the intricate internal structure.